\documentclass[a4paper]{report}
\usepackage[utf8]{inputenc}
\usepackage[T1]{fontenc}
\usepackage{RJournal}
\usepackage{amsmath,amssymb,array}
\usepackage{booktabs}
\usepackage{float}
\usepackage{xspace}
\usepackage{xcolor}
\usepackage{listings} 
\usepackage{enumitem} 
\usepackage{calc} 

\definecolor{mygreen}{rgb}{0,.7,0}
\definecolor{mygray}{rgb}{0.4,0.4,0.4}
\definecolor{mymauve}{rgb}{0.58,0,0.82}

\newcommand{\cpp}{\mbox{C/C++}\xspace}
\newcommand{\cfss}{compiled functions\slash subroutines\xspace}
\newcommand{\cfs}{compiled function\slash subroutine\xspace}
\newcommand{\dc}{\code{.C64()}\xspace}

\newcommand{\expressions}{\code{expressions}\xspace}
\newcommand{\lists}{\code{lists}\xspace}
\newcommand{\character}{\code{character}\xspace}
\newcommand{\complex}{\code{complex}\xspace}
\newcommand{\raw}{\code{raw}\xspace}
\newcommand{\logical}{\code{logical}\xspace}
\newcommand{\numeric}{\code{numeric}\xspace}
\newcommand{\integer}{\code{integer}\xspace}
\newcommand{\integers}{\code{integers}\xspace}
\renewcommand{\int}{\code{int}\xspace}

\newcommand{\double}{\code{double}\xspace}
\newcommand{\doubles}{\code{doubles}\xspace}
\newcommand{\shortint}{\code{int32\_t}\xspace}
\newcommand{\longint}{\code{int64\_t}\xspace}
\newcommand{\Rxlen}{\code{R\_xlen\_t}\xspace}
\newcommand{\longvector}{long vector\xspace}

\newcommand{\longvectors}{long vectors\xspace}

\newcommand{\ffi}{foreign function interface\xspace}
\newcommand{\mic}{modern interfaces to \cpp code\xspace}
\newcommand{\BITint}{\mbox{64-bit} integer\xspace}
\newcommand{\BITints}{\mbox{64-bit} integers\xspace}
\newcommand{\BIT}{\mbox{64-bit}\xspace}
\newcommand{\bitint}{\mbox{32-bit} integer\xspace}
\newcommand{\bitints}{\mbox{32-bit} integers\xspace}

\newcommand{\eg}{\mbox{e.\,g.\,}\xspace}
\newcommand{\ie}{\mbox{i.\,e.\,}\xspace}
\renewcommand{\R}{R\xspace}

\makeatletter
\DeclareRobustCommand*\textsubscript[1]{%
  \@textsubscript{\selectfont#1}}
\def\@textsubscript#1{%
  {\m@th\ensuremath{_{\mbox{\fontsize\sf@size\z@#1}}}}}
\makeatother

\graphicspath{{Figures/}}

\begin{document}

\sectionhead{Contributed research article}
\volume{XX}
\volnumber{YY}
\year{20ZZ}
\month{AAAA}

\begin{article}


\title{\pkg{dotCall64}: An Efficient Interface to Compiled \cpp and Fortran Code Supporting Long Vectors}
\author{by Florian Gerber, Kaspar M\"{o}singer and Reinhard Furrer}

\maketitle

\abstract{ 
The \R~functions \code{.C()} and \code{.Fortran()} can be used to call compiled \cpp and Fortran code from \R. 
This so-called \ffi is convenient, since it does not require any interactions with the C API of \R.
However, it does not support \longvectors (\ie, vectors of more than $2^{31}$ elements).
To overcome this limitation, the \R~package \pkg{dotCall64} provides \dc,
which can be used to call compiled \cpp and Fortran functions.
It transparently supports \longvectors and does the necessary castings to pass \numeric \R vectors to \BITint arguments of the compiled code.
Moreover, \dc features a mechanism to avoid unnecessary copies of function arguments,
making it efficient in terms of speed and memory usage. 
}

\section{Introduction}
The interpreted character of \R makes it a convenient front-end for a wide range of applications. 
Although \R provides a rich infrastructure, it can be advantageous to extend \R programs with compiled code written in \cpp or Fortran \citep{Euba:Kupr:11}. 
According to \cite{Cham:08}, reasons for such an extension are 
the access to new and trusted computations, the increase in computational speed, and the object referencing capabilities. 
For completeness, we also list the reasons against such an extension, which include
an increased workload to write, maintain, and debug the software, platform dependencies, and a less readable source code.

\R provides two types of interfaces to call compiled code documented in ``Writing \R Extensions'' \citep{r_ext}.
First, the \dfn{\mic} feature the \R functions \code{.Call()} and \code{.External()}. 
It enables accessing, modifying, and returning \R objects from \cpp using the C API of \R \citep{Wick:11}. 
On one hand, this is convenient when the \cpp code is specifically written to be used with \R.
In that case, the C API serves as a glue between \R and \cpp, providing some \R functionality and control over copying \R objects on the \cpp level. 
On the other hand, it requires the user to learn the C API of \R.
Especially, when an \R interface is built on top of existing \cpp code this constitutes an additional effort.
Since \R has no Fortran API, the \mic are not suitable to embed Fortran code into \R.
Second, the \dfn{\ffi} provides the \R functions \code{.C()} and \code{.Fortran()}.
This interface allows the compiled code to read and modify atomic \R vectors, 
which are exposed as the corresponding \cpp and Fortran types, respectively.
Thus, no additional API is required, making it favorable for embedding \cpp and Fortran code that is not specifically designed for \R.

On top of these interfaces provided by \R, \R~packages exist that simplify the integration of compiled code into \R.
One such \R~package is \CRANpkg{inline} \citep{inline}, which allows the user to dynamically define \R\ functions and S4 methods with inlined compiled code. 
Other examples are \CRANpkg{Rcpp} \citep{Rcpp0,Rcpp1,Rcpp2} and its extensions \CRANpkg{RcppArmadillo} \citep{RcppArmadillo0,RcppArmadillo1}, \CRANpkg{RcppEigen} \citep{RcppEigen0,RcppEigen1}, \CRANpkg{RcppParallel} \citep{RcppParallel0}, and
\CRANpkg{Rcpp11} \citep{Rcpp11}, which greatly simplify the extension of \R with C++ code. 
Similar to the \mic, the \pkg{Rcpp} package family is designed to extend \R with compiled code that is specifically written for that purpose. 

\begin{figure}
\centering
\includegraphics[trim = 4mm 14mm 4mm 12mm, clip, width=.7\textwidth]{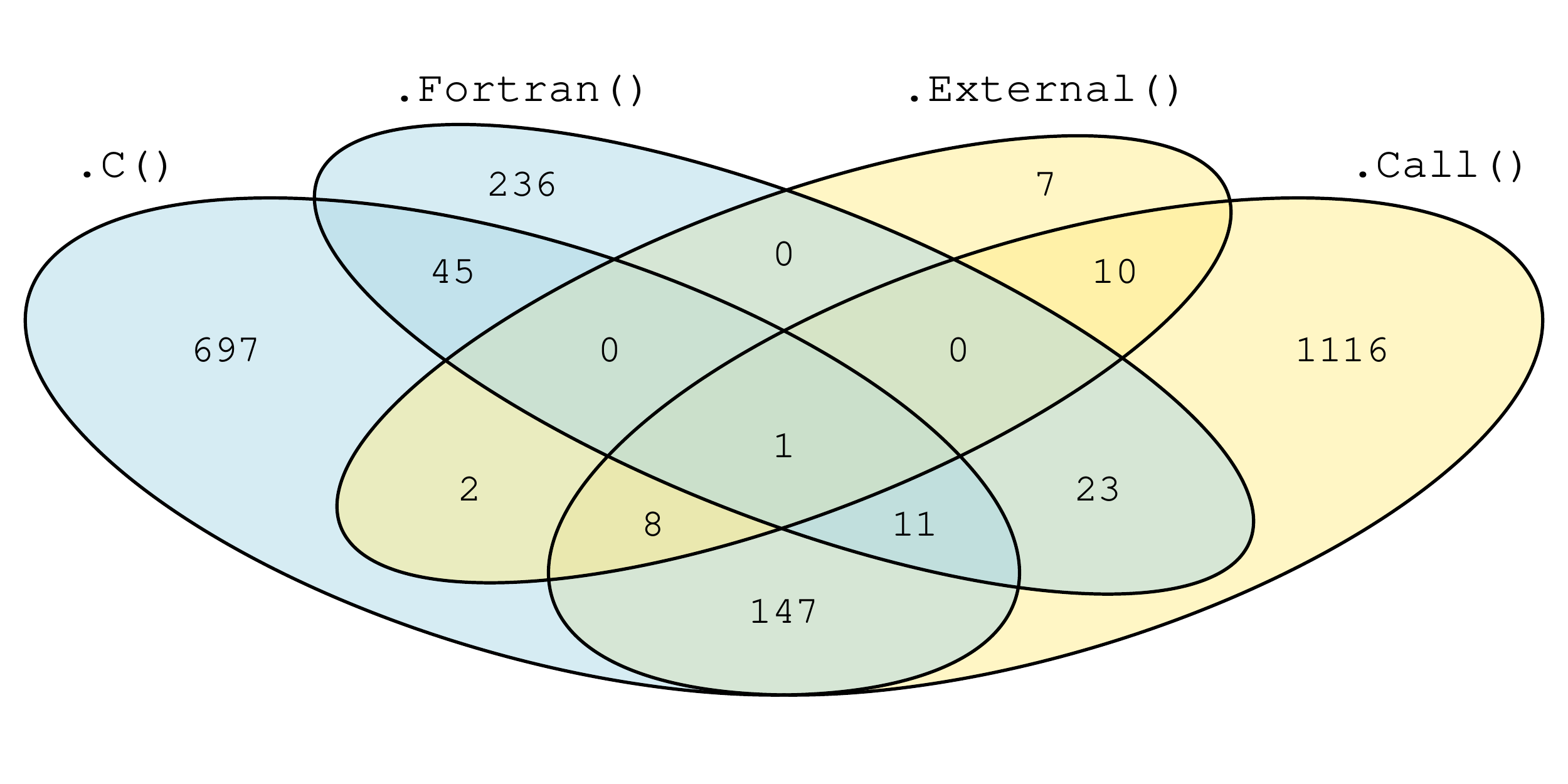} 
\caption{Number of \R\ packages on CRAN using the indicated \R~functions to interface \cpp or Fortran code (as of 2016-09-02). 
Note that \R~packages linking to \pkg{Rcpp} use \code{.Call()}.}
\label{fig:venn}
\end{figure}

Building \R packages is a way to share compiled code across different platforms. 
(See, \eg, \citealp{Plum:11} for comments on including portable C++ code in \R packages.)
As of \mbox{09-02-2016}, $2'303$ of the $9'079$ \R~packages on CRAN (\url{http://www.cran.r-project.org/}) include compiled \cpp and\slash or Fortran code using both the \ffi and the \mic with a similar frequency.
Figures~\ref{fig:venn} gives an overview of the number of packages using \code{.C()}, \code{.Fortran()}, \code{.Call()}, and \code{.External()}. 

In the remainder of this article, we focus on the intention to embed compiled code into \R without using its C API.
An example of an \R~package using that type of interface is the SPArse Matrix package \CRANpkg{spam} \citep{spam,Furr:13,Gerb:Furr:15}, which is built around the Fortran library SPARSKIT \citep{Saad:94}.
Here, the \R function \code{.Fortran()} from the \ffi seems to be suitable.
Conversely, using the \mic is also possible but requires adding an additional layer of C code to enable communication between \R and the compiled Fortran code. 
However, using \code{.Fortran()} is also not satisfying, since it lacks flexibility and performance, as also stated in its help page:
``These functions [\code{.C()} and \code{.Fortran()}] can be used to make calls to compiled C and Fortran 77 code. 
Later interfaces are ‘.Call’ and ‘.External’ which are more flexible and have better performance.''
Two of the missing features of the \ffi are:
\begin{itemize}
\item support of \longvectors,
\item a mechanism to avoid unnecessary copies of \R vectors.
\end{itemize}
The latter is the reason for the lower performance of the \ffi compared to the \mic. 
Since the \ffi does not allow \R vectors to be passed to compiled code by reference (without copying),
it is especially impractical for big data application.
The missing features of the \ffi motivated the development of the \R~package \CRANpkg{dotCall64} \citep*{dotCall64}, which is presented in this article.

\section{Limitations of the \ffi}
To set the scene for \pkg{dotCall64}, we first discuss some limitations of the \ffi and give insights into the \R implementation of \longvectors. 

\subsection{Long vectors}
The \ffi does not support \longvectors; see \code{help("long vector")}. 
To understand why extending it to support \longvectors is a non-trivial task, 
we give more details on the \longvector implementation of \R.
In \R, vectors are one of the most basic object types underlying more complex objects, such as matrices and arrays. 
They can be thought of as strings of elements that can be indexed according to their relative positions. 
Prior to version 3.0.0, the length of vectors was limited to $2^{31}-1$ elements and indexing thereof was exclusively based on \R vectors of type integer.
More precisely, the latter are signed \bitint vectors having a value range of $[-2^{31}+1,\,2^{31}-1]$. 
Starting from the release of version 3.0.0 in early 2013, support for so-called \longvectors was supplied. 
That is, atomic (\raw, \logical, \integer, \numeric, \complex, and \character) vectors, \lists, and \expressions can now have up to $2^{52}$ elements. 
The introduction of \longvectors was done with minimal changes in \R and especially, without changing or adding a \BITint data type. 
Vectors of lengths less than $2^{31}-1$ remain unchanged and addressing elements thereof still uses \R vectors of type \integer.
In contrast, \longvectors use \numeric vectors of type \doubles to address elements, which are integer precise up to $2^{52}$.
This implied changes in some \R functions, such as \code{length()}, which returns an \integer or a \double type depending on whether the input vector is a \longvector.
\begin{example} 
> typeof(length(integer(1)))
[1] "integer"
> typeof(length(integer(2^31)))
[1] "double"
\end{example}
Note that \code{as.numeric()} returns a \double type and \code{as.integer()} returns an \integer type, 
though both \integer and \double type are of class \code{"numeric"},
see the ``Note on names'' section in the \R help page \code{help("is.double")}. 

While the \R implementation of \longvectors favors backwards compatibility,
care is needed when manipulating those with compiled code. 
We distinguish between passing \longvectors and indexing \longvectors:
The former requires passing vectors of more than $2^{31}-1$ elements to complied code and is trivial. 
The latter is challenging, since the indexing \R vector is of type \double, whereas the compiled code would naturally expect a \BITint type. 
To overcome this discrepancy, one needs to cast the indexing vector from a \double to a \BITint type before calling the compiled code and back-cast it afterwards.

\paragraph{Technical note:}
This section gives technical insights into the underlying C implementation of \longvectors in \R and may be skipped without loss of the general idea. 
We refer to the source code of \R version~3.3.1 in several places and show relevant parts thereof in the appendix.
Information on the current and future directions of \longvectors and \BIT types in \R can be found in ``R Internals'' \citep[][Section~12]{r_int}.

In \R, vectors are made out of a header of type \code{VECSEXP} that is followed by the actual data (Listing~\ref{list:Rinternals}, line 272).  
The header contains a field \code{length} of type \code{R\_len\_t}, which is defined as signed \shortint (a \bitint). 
Thus, that \code{length} field cannot capture the length of a \longvector.  
Instead, it is set to \code{-1} whenever the length of the vector is larger than $2^{31} - 1$, 
and an additional header of type \code{R\_long\_vec\_hdr\_t} is prefixed.
The prefixed header has a field \code{length} of type \Rxlen, which is defined as \code{ptrdiff\_t} type 
(Listing~\ref{list:Rinternals}, line 75) being ``[...] the signed \integer type of the result of subtracting two pointers. 
This will probably be one of the standard signed \integer types (short int, int or long int), but might be a nonstandard type that exists only for this purpose'' \citep[][Appendix~A.4]{gnu_c_lib}.   

This implementation has the advantage that the existing code does not need to be changed and still works with vectors having less than $2^{31}$ elements.
Hence, the C code of \R can be changed successively to support \longvectors throughout several \R versions, as opposed to changing the entire C code in one step.   
To make C code compatible with \longvectors, adaptations are needed. 
For example, the widely used C function \code{R\_len\_t length(SEXP s)} (Listing~\ref{list:Rinlinedfuns}, line 124) returns the length of a \code{SEXP} (S expression) as a \code{R\_len\_t}.
Thus, all instances of that function have to be replaced with calls to the \BIT counterpart (\ie, the function \code{R\_xlen\_t xlength(SEXP s)} given in line 159 of Listing~\ref{list:Rinlinedfuns}).

\subsection{Copying arguments}
The \ffi exposes pointers to \R vectors to compiled code.
In order to avoid any corruption of \R vectors, they are copied and the compiled code receives pointers to copies of the \R vectors. 
One exception is when the \R vector has the named status 0 (\ie, the object is not bound to any symbol); see ``Writing \R Extensions'' \citep[][Section~5.9.10]{r_ext}.
This is the case when the passed \R vector is an evaluated constructor (\eg, \code{integer(1)}).
This is often used when the only purpose of the \R vector is to capture results from the compiled code.

Another situation in which there is no need for copying \R vectors is when the compiled code only reads an \R vector without modifying it. 
However, the \ffi does not allow the user to avoid copying of \R vectors (with named status 1 or 2), which leads to a significant computational overhead, especially for large vectors.
Note that prior to \R version~3.2.0, the copying of \R vectors could be avoided by setting the argument \code{DUP} of \code{.C()} and \code{.Fortran()} to \code{FALSE}.
In later \R versions, this argument is depreciated and users are referred to the \mic as a more flexible interface; see \code{help(".C")} and ``R NEWS'' \citep[][]{r_news}.

\section[The R package dotCall64]{The \R~package \pkg{dotCall64}}
The limitations of the \ffi discussed above have motivated the development of the \R~package \pkg{dotCall64}.
Its main function is \dc, which can be used to interface compiled code.
In contrast to \code{.C()} and \code{.Fortran()}, 
it supports \longvectors and \BITint arguments of complied \cfss and provides a mechanism to control duplication of function arguments. 
Emphasis was put on providing a trustworthy implementation featuring 
structured \R and C source code, documentation, examples,
unit tests implemented with \CRANpkg{testthat} \citep{testthat}, 
and \R scripts containing the later presented performance measurements.

\subsection{Usage of the \R  function \dc}
The function \dc can be used as an enhanced replacement of the \ffi and is equally easy to use;
see also the documentation in the reference manual \citep*{dotCall64}.
Its syntax resembles that of the function \code{.C()}, and both functions have common arguments as shown in Table~\ref{tab:args}.
\begin{table}[h]
\vspace*{-6mm}
\centering
{\small
\begin{tabular}{llll}
\toprule
\multicolumn{2}{c}{\code{.C()}}&\multicolumn{2}{c}{\dc}\\
\cmidrule(r){1-2} \cmidrule(r){3-4}  
\code{arguments}  & \code{defaults}                       & \code{arguments}   & \code{defaults}             \\ 
  \midrule
\code{.NAME}       &              &  \code{.NAME}      &                                                     \\
                   &              &  \code{SIGNATURE}  &                                                     \\
\samp{...}         &              &  \samp{...}        &                                                     \\
                   &              &  \code{INTENT}     & \code{NULL}                                         \\
\code{NAOK}        & \code{FALSE} &  \code{NAOK}       & \code{FALSE}                                        \\
$^*$\code{DUP}     & \code{TRUE}  &                    &                                                     \\
\code{PACKAGE}     &              &  \code{PACKAGE}    & \code{""}                                           \\
$^*$\code{ENCODING}&              &                    &                                                     \\
                   &              &  \code{VERBOSE}    & \code{getOption("dotCall64.verbose")}               \\
\bottomrule 
\end{tabular}
}
\vspace*{-1mm}
\caption{
Arguments and default values of the \R function \code{.C()} from the \ffi and \dc from \pkg{dotCall64}.
The depreciated arguments of \code{.C()} are marked with ``$^*$''.
}
\label{tab:args}
\end{table}
\newpage
\begin{table}[h]
\vspace*{-1mm}
\centering
{\small
\begin{tabular}{lllll}
\toprule
\code{SIGNATURE}   & \cpp type      & Fortran type            & \R type       & cast \\ 
  \midrule
\code{"double"}    & \code{double}   & \code{double precision} & \code{double} & no   \\
\code{"int"}       & \code{int}      & \code{integer (kind = 4)}        & \code{integer}& no   \\
\code{"int64"}     & \code{int64\_t} & \code{integer (kind = 8)}        & \code{double} & yes  \\
\bottomrule 
\end{tabular}
}
\vspace*{-1mm}
\caption{
Supported \code{SIGNATURE} arguments of \dc and the corresponding \cpp, Fortran, and \R data types.  
The column ``cast'' indicates whether casting is necessary.   
}
\label{tab:types}
\end{table}
The required arguments of \dc are:
\begin{itemize}[leftmargin=1.7cm]
\item[\code{.NAME}] The name of the compiled \cpp function or Fortran subroutine. 
\item[\code{...}] Up to $65$ \R vectors to be accessed by the compiled code.
\item[\code{SIGNATURE}] A character vector of the same length as the number of arguments of the \cfs. 
Each string specifies the signature of one such argument. 
Accepted signatures are \code{"integer"}, \code{"double"}, and \code{"int64"}. 
The \R, \cpp, and Fortran types corresponding to these specifications are given in Table~\ref{tab:types}.
\end{itemize}
With that, the following call to the compiled C function \code{void get\_c(double input,\ int index,\ double output)} using \code{.C()} can be replaced by its \dc counterpart. 
Therefore, for example, 
\begin{example}
> .C("get_c", input = as.double(1:10), index = as.integer(9), output = double(1))
\end{example}
becomes
\begin{example}
> .C64("get_c", SIGNATURE = c("double", "integer", "double"), 
+      input = 1:10, index = 9, output = 0)
\end{example}
While more detailed code examples are given later, 
this is enough to highlight some features of \dc.
First, \dc does require the additional argument \code{SIGNATURE} specifying the argument types of the \cfs. 
In return, it coerces the provided \R vectors to the specified signatures making the \code{as.double()} and \code{as.integer()} statements unnecessary. 
Second, all provided arguments can be \longvectors.
Third, if one of the arguments of the compiled function is a \BITint 
(\code{int64\_t} in the case of \cpp functions, and \code{integer (kind = 8)} types for Fortran subroutines), it is enough to set the corresponding \code{SIGNATURE} argument to \code{"int64"} to successfully evaluate the function.  
That is, \dc does the necessary \double to \BITint and \BITint to \double castings before and after evaluating the compiled code, respectively.

Additional arguments of \dc are the following:
\begin{itemize}[leftmargin=1.7cm]
\item[\code{INTENT}] A character vector of the same length as the number of arguments of the \cfs. 
Each string specifies the intent of one such argument. 
Accepted intents are \code{"rw"} (read and write), \code{"r"} (read), and \code{"w"} (write). 
\item[\code{NAOK}] A logical flag specifying whether the \R vectors passed though \samp{...} are checked for missing and infinite values. 
\item[\code{PACKAGE}] A character vector of length one restricting the search path of the \cfs to the specified package. 
\item[\code{VERBOSE}] If \code{0} (default), no warnings are printed. If \code{1} and \code{2}, then warnings for tuning and debugging purposes are printed. 
\end{itemize}
A complete list of arguments including their default values is also given in Table~\ref{tab:args}.

The argument \code{INTENT} influences the copying of \R vectors and can be seen as an enhanced version of the depreciated \code{DUP} argument of \code{.C()}.
By default, all intents are set to ``read and write'' implying that the compiled code receives pointers to copies of the \R vector given to \samp{...}.
This behavior is desirable when the compiled function reads the corresponding \R vectors and modifies (writes to) them.
For arguments of the \cfs that are only read and not modified, the intent can be set to ``read.''
With that, the compiled code receives pointers to the corresponding \R vectors itself.
While this avoids copying, it is absolutely necessary that the compiled code does not alter these vectors, 
as this corrupts the corresponding \R vectors in the current \R session. 
For arguments that are only used to write results into it, the intent ``write'' is suitable. 
To obtain the desired performance gain, the corresponding \R vectors passed to \samp{...} have to be of class \code{"vector\_dc"}.
\R objects of that class contain information on the type and length of the vectors.
They can be constructed with the \R function \code{vector\_dc()}, taking the same arguments as \code{vector()} from the \pkg{base} \R~package. 
For example, instead of passing the \R vector \code{vector(mode = "numeric",} \code{length = 8)},
the following \R object should be passed. 
\begin{example}
> vector_dc(mode = "numeric", length = 8)
$mode
[1] "numeric"

$length
[1] 8

attr(,"class")
[1] "vector_dc" "list"     
\end{example}
Based on this information, \dc allocates the corresponding vector (initialized with zeros). 
That vector is then exposed to the compiled function to write into it.
Note that specifying the suitable intent may reduce computation time by avoiding unnecessary copying of \R vectors
and by avoiding unnecessary \double to \BIT \integer and \BIT \integer to \double castings for \code{SIGNATURE} \code{= "int64"} type arguments. 
More details on the other arguments are given in the package manual of \code{dotCall64} \citep*{dotCall64}.

\subsection{Implementation of the \R function \dc}
\begin{figure}[tb]
\centering
\includegraphics[trim = 142mm 15mm 95mm 40mm, clip, width=.7\textwidth]{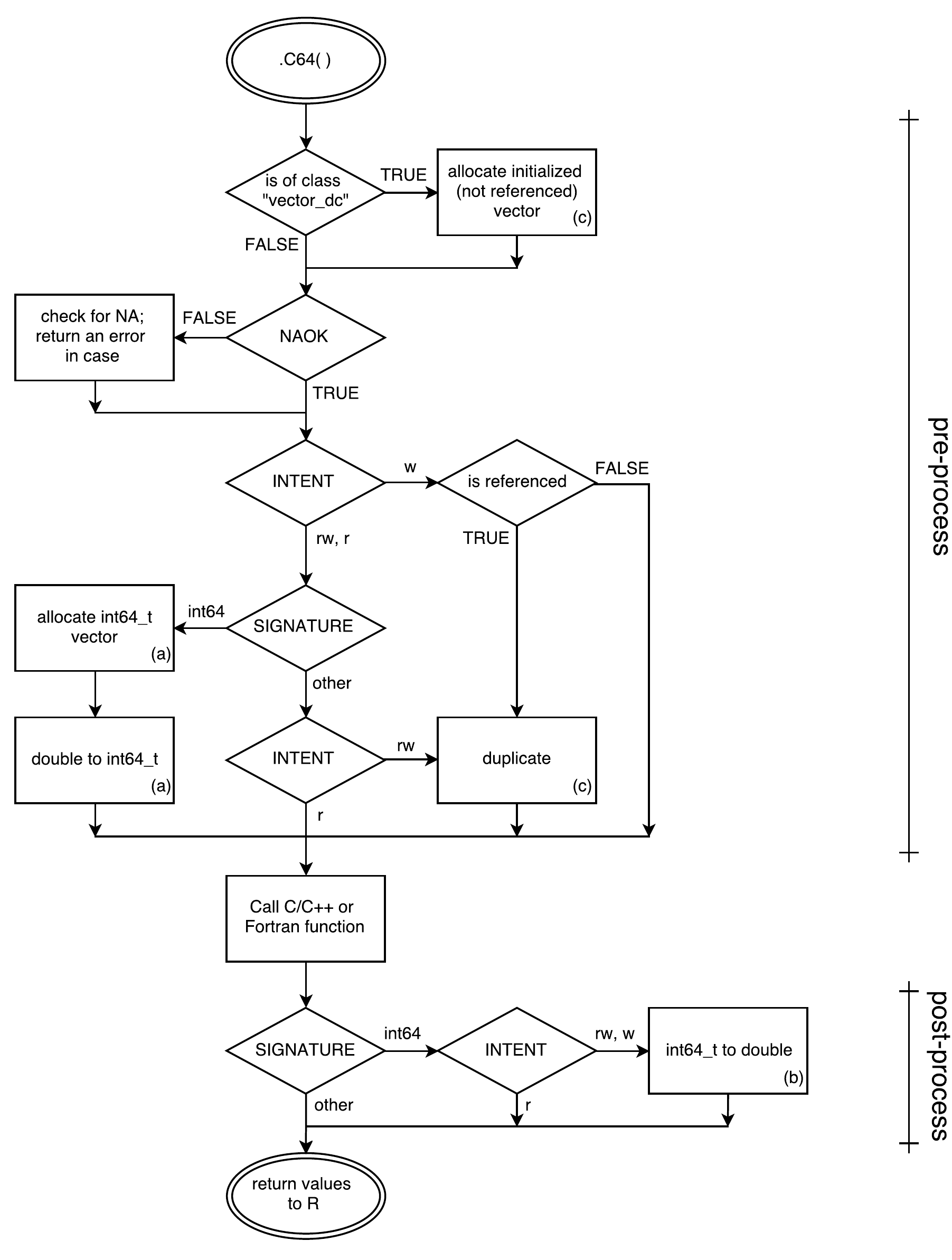} 
\caption{Flowchart of the involved processes when using \dc to call a \cfs with one argument. 
In the pre-process phase, the provided \R vector passed through \samp{...} is checked and 
prepared according to the arguments \code{NAOK}, \code{SIGNATURE}, and \code{INTENT}. 
Then, the \cfs specified with the argument \code{.NAME} is called. 
Finally, the vector is back-cast in the post-process phase if necessary. 
} 
\label{fig:flow}
\end{figure}

The function \dc uses the function \code{.External()} from the \mic to directly pass all provided arguments to the C function \code{dC64()}.
After basic checks of the provided arguments, the function proceeds as schematized in Figure~\ref{fig:flow}.
Note that the flowchart depicts the procedure for the case in which the \cfs has only one argument. 
Otherwise, \code{dC64()} repeats the depicted scheme for all arguments. 

One aspect to highlight is the castings of \R vectors for \code{SIGNATURE} \code{= "int64"} arguments. 
For such arguments, the \double to \longint casting is done for the intents ``read and write'' and ``read'';
see the boxes labeled with~(a).
In that case, duplication is not necessary, as the implemented casting allocates a new vector anyway.  
The back-casting from \longint to \double is only done for the intents ``read and write'' and ``write'';
see the box labeled with~(b).

Moreover, an argument of \code{SIGNATURE} different from \code{"int64"} with intent ``read and write'' is duplicated in any case; see boxes labeled with~(c).
If the intent is ``read,'' it is not duplicated, 
and if the intent is ``write,'' the argument is only duplicated when it has a reference status different from 0.
\R vectors increase their reference status when they are passed to an \R function, and 
therefore a safe way to allocate a zero initialized vector without copying is to pass an \R object of class \code{"vector\_dc"}.  

As casting is an expensive operation in terms of computational time, we distribute this task to multiple threads using openMP, if available \citep{Dagu:Meno:98,openMP}.
Note that the number of used threads can be controlled with the \R function \code{omp\_set\_num\_threads()} from the package \CRANpkg{OpenMPController} \citep{OpenMPController}.
The package \pkg{dotCall64} can also be compiled without the openMP feature by removing the flag \samp{\$(SHLIB\_OPENMP\_CFLAGS)} in the \file{src/Makevars} file of the source code.

\section{Examples}\label{sec:example}
We showcase the function \dc from the \R~package \pkg{dotCall64} with an example function implemented in C and Fortran.
Besides the calls thereof via \dc, the C and Fortran function definitions 
and the commands to compile and load the code are given.
A direct comparison with \code{.C()} shows the limitations of the \ffi and 
that it is straight forward to overcome these with \dc. 
Moreover, the similarities and differences in the syntax become visible. 
The considered example function takes the arguments \samp{input} (\double), \samp{index} (\integer), and \samp{output} (\double) 
and writes the element of \samp{input} at the position specified with \samp{index} to \samp{output}.

\subsection{Interface \cpp code}
A C implementation of the described example function is given next.
\begin{example}
void get_c(double *input, int *index, double *output) {
    output[0] = input[index[0] - 1];
}
\end{example}
We write the function into \file{get\_c.c} and compile it with the command line command \samp{R CMD SHLIB get\_c.c}.
The resulting dynamic shared object (\file{get\_c.so} on our Linux platform) must be loaded into \R before 
the compiled function can be called.
Note that, in the following \R code, the extension of the shared object is replaced with \code{.Platform\$dynlib.ext} to make the code platform independent.
\begin{example}
> dyn.load(paste0("get_c", .Platform$dynlib.ext))
\end{example} 
One can use the \ffi to call this function. 
We use the \R functions \code{as.double()} and \code{as.integer()} to ensure that the types of the passed \R vectors
match the signature of the C function \code{get\_c()}. 
\begin{example}
> .C("get_c", input = as.double(1:10), index = as.integer(9), output = double(1))$output
[1] 9
\end{example} 
Next, we try to use the same call with a \longvector \code{x\_long} passed to the argument \samp{input} of \code{get\_c()}.
\begin{example} 
> x_long <- double(2^31); x_long[9] <- 9; x_long[2^31] <- -1
> .C("get_c", 
+    input = as.double(x_long), index = as.integer(9), output = double(1))$output
Error: long vectors (argument 1) are not supported in .Fortran
\end{example}
As expected, \code{.C()} throws an error because it does not support \longvectors. 
The error---and the confusing error message referring to \code{.Fortran()} instead of \code{.C()}---can 
be avoided by replacing \code{.C()} with \dc.
This allows the evaluation of the C function \code{get\_c()} with the \longvector \code{x\_long}.
Additionally, \dc requires the argument \code{SIGNATURE} encoding the signatures of the arguments of \code{get\_c()}. 
This information is used to coerce all provided \R vectors to the specified signatures.
Thus, it is no longer necessary to reassure that the types of the passed \R vectors
match the signature of the compiled function.
\begin{example} 
> install.packages("dotCall64")
> library("dotCall64")
> .C64("get_c", SIGNATURE = c("double", "integer", "double"),
+      input = x_long, index = 9, output = double(1))$output
[1] 9
\end{example}
In contrast to the call using \code{.C()}, the ninth element of the \longvector \code{x\_long} is returned. 
However, the argument \samp{index} of \code{get\_c()} is of type \int (a \bitint), 
and hence, elements at positions beyond $2^{31}-1$ cannot be extracted.
To overcome this, we adapt the definition of the C function \code{get\_c()} and replace  
the \int type in the declaration of the argument \samp{index} with the \longint type, 
which is defined in the C header file \file{stdint.h}. 
\begin{example}
#include <stdint.h>
void get64_c(double *input, int64_t *index, double *output) {
    output[0] = input[index[0] - 1];
}
\end{example}
We write the function into \file{get64\_c.c} and compile it with \samp{R CMD SHLIB get64\_c.c} to obtain the dynamic shared object (\file{get64\_c.so} on our platform).
Because of the \longint argument, it is not possible to call this function with \code{.C()}. 
On the other hand, \dc can interface this function 
when the second element of the \code{SIGNATURE} argument is set to \code{"int64"}.
\begin{example}
> dyn.load(paste0("get64_c", .Platform$dynlib.ext))
> .C64("get64_c", SIGNATURE = c("double", "int64", "double"),
+      input = x_long, index = 2^31, output = double(1))$output
[1] -1
\end{example}
In the call above, the function \dc casts the argument \samp{index} from \double (the \R representation of \BITints) into a \longint type vector before calling \code{get64\_c()}, 
and back-casts it from \longint to \double afterwards. 

\subsection{Interface Fortran code} 
The function \dc can also be used to interface compiled Fortran code. 
To highlight some Fortran specific features, we translate the C function \code{get\_c()} into the Fortran subroutine \code{get\_f()}.
\begin{example} 
      subroutine get_f(input, index, output)
      double precision :: input(*), output(*)
      integer :: index    
      output(1) = input(index)
      end
\end{example} 
Note that we only use lower case letters in the Fortran function and variable names to avoid unnecessary symbol-name translations.
We write the function into the \file{get\_f.f} and compile it with \samp{R CMD SHLIB get\_f.f} to obtain the dynamic shared object (\file{get\_f.so} on our platform).
In contrast to \code{.Fortran()}, \dc allows passing pointers to \longvectors. 
\begin{example}
> dyn.load(paste0("get_f", .Platform$dynlib.ext))
> .C64("get_f", SIGNATURE = c("double", "integer", "double"),
+      input = x_long, index = 9, output = double(1))$output
[1] 9
\end{example}
Again, elements with positions beyond $2^{31}-1$ cannot be accessed, 
since the argument \samp{index} is of type \integer and compiled as a \bitint by default. 
To make \code{get\_f()} compatible with \BITints, we can either change the declaration of \samp{index} to \samp{integer (kind = 8) index} in \file{get\_f.f} or 
leave the Fortran code unchanged and set the following compiler flag to compile \integers as \BITints.
\begin{example}
MAKEFLAGS="PKG_FFLAGS=-fdefault-integer-8" R CMD SHLIB get_f.f 
\end{example}
Note that both the \samp{kind = 8} declaration and the \samp{-fdefault-integer-8} flag are valid for the GFortran compiler \citep{gfortran} and may not have the intended effect using other compilers.
The resulting dynamic shared object from the command above (\file{get\_f.so} on our platform) can be called from \R as follows.
\begin{example}
> dyn.load(paste0("get_f", .Platform$dynlib.ext))
> .C64("get_f", SIGNATURE = c("double", "int64", "double"),
+      input = x_long, index = 2^31, output = double(1))$output
[1] -1
\end{example}   

\subsection[Extend R packages to support long vectors]{Extend \R packages to support long vectors}
Extending \R packages to support \longvectors allows developers to distribute compiled code featuring \BITints with an \R user interface.
Given the popularity of \R, this is a promising approach to make such software available to many users.
With the function \dc, the workload of extending an \R package to support \longvectors is reduced to the following tasks:
\begin{itemize}
\item replace the R function to call compiled code with \dc,
\item replace the \bitint type declarations in the compiled code with a \BITint declaration.
\end{itemize}
The latter task implies replacing all \int type declarations in \cpp code with \code{int64\_t} type declarations
and replacing all \integer type declarations in Fortran code with \samp{integer (kind = 8)}. 
In both cases, the replacements can be automatized (\eg, with the stream editor \citealp{sed}).
If the considered Fortran code does not explicitly declare the bits of the integers, 
an alternative approach is to set the compiler flag \samp{-fdefault-integer-8} to compile integers as \BITints using GFortran compilers. 
This is convenient because the Fortran code does not need to be changed at all in that case. 

A more elaborate extension could feature two versions of the compiled code:
one with \bitints and the other one with \BITints. 
Then, the \R function can dispatch to either version according to the sizes of the involved vectors. 
This avoids \double to \BITint castings when only vectors with less than $2^{31}-1$ elements are involved. 
It is convenient to manage two versions of compiled code by putting them into two separate R packages. 
The first package includes the compiled code with \bitints together with the \R code and the documentation. 
This package can be used independently as long as no \longvectors are involved. 
The second package can be seen as an add-on package and includes only the compiled code with integers declared as \BITints. 
Thus, loading both packages enables \longvector support. 
This separation into two packages has the advantage that the compiled functions featuring \bitints and their \BIT counterparts can have the same name.
The desired function is then specified by setting the appropriate \code{PACKAGE} argument of \dc.

As a proof of concept, we extended the sparse matrix algebra \R~package \pkg{spam} to handle 
sparse matrices with more the $2^{31}-1$ non-zero elements. 
From the user perspective, the syntax to manipulate such matrices remains the same.
In fact, \pkg{spam} users may not even notice the extension.
In the case, in which the number of non-zero entries of a matrix exceeds $2^{31}-1$ and 
the add-on package \pkg{spam64} is loaded,
\pkg{spam} automatically dispatches to the compiled code with \BITints. 
The new capabilities of \pkg{spam} and \pkg{spam64} were illustrated with a parametric model of a non-stationary spatial covariance matrix fitted to satellite data.
More information on \pkg{spam64} and the data example is given by \cite*{Gerb:Mosi:Furr:16}.

\section{Performance}\label{sec:performance}
There are different settings in which the elapsed time to interface compiled code is relevant.
One of those is when the compiled code is interfaced often and takes only a short time to evaluate. 
Here, the overhead of the interface becomes relevant, which is in the order of a few microseconds for \dc.
Another such setting is when large and possibly \longvectors are passed through \dc.
In that case, the overhead is negligible, as other services of the interface and the execution of the compiled code take up several orders of magnitude more time.
When \dc is used to interface \BITint arguments of the compiled code, 
the largest share of the elapsed time is caused by the \double to \BITint and \BITint to \double castings. 
Since castings are implemented with openMP, the elapsed time thereof also depends on the number of used threads. 
Besides that, copying objects and checking them for missing\slash infinite values are also time-consuming operations. 

Another performance aspect is peak memory usage. 
Using the default arguments of \dc, its peak memory usage is about twice the size of the \R vectors passed through \samp{...}, and hence, is similar to \code{.C()}. 
An exception where the peak memory usage is reduced is indicated below. 

\subsection[Performance relevant arguments of .C64()]{Performance relevant arguments of \dc}
Further, \dc provides arguments to optimize calls to compiled code,
one of which is the argument \code{INTENT}, which is set to ``read and write'' by default.
Since many \cfss only read or write to certain arguments, it is safe to avoid copying in some cases. 
For example, the C function \code{get64\_c()}, as defined above, only reads the arguments \samp{input} and \samp{index} and only writes to the argument \samp{output}.
Thus, we can set the \code{INTENT} argument of \dc to \code{c("r", "r", "w")} and pass the argument with intent ``write'' as objects of class \code{"vector\_dc"} to reduce the copying of \R vectors to a minimum.
Another significant performance gain is obtained by setting the argument \code{NAOK} to \code{TRUE}.
This avoids checking the \R vectors passed through \samp{...} for \code{NA}, \code{NaN}, and \code{Inf} values. 
Small-scale performance gains can be achieved by setting the \code{PACKAGE} argument, which reduces the time to find the compiled code, and by setting \code{VERBOSE} \code{= 0}, which avoids the execution of \samp{getOptions("dotCall64.verbose")}. 
Similar speed considerations that are partially applicable to \dc are given in ``Writing \R Extensions'' \citep[][Section 5.4.1]{r_ext}.
An optimized version of the call to the C function \code{get64\_c()}, taking the discussed performance considerations into account, is given next.
\begin{example}
> .C64("get64_c", SIGNATURE = c("double", "int64", "double"),
+      input = x_long, index = 2^31, output = numeric_dc(1),
+      INTENT = c("r", "r", "w"), NAOK = TRUE, PACKAGE = "dotCall64", VERBOSE = 0)
\end{example} 

\subsection{Timing measurements}
In the following, we present detailed timing measurements and benchmark \dc against \code{.C()}, where possible. 
We consider the following C function contained in the \R package \pkg{dotCall64}.
\begin{example} 
void BENCHMARK(void *a) { }
\end{example} 
This function takes one pointer \samp{a} to a variable of an unspecified data type and does no operations with it. 
Thus, the elapsed time to call \code{BENCHMARK()} from \R is dominated by the performance of the used interface.
We measure the time to call this function with different \code{NAOK} and \code{INTENT} settings of \dc and benchmark it against \code{.C()} using \CRANpkg{microbenchmark} \citep{microbenchmark}. 
To get an estimate of the measurement uncertainty, we repeated the measurements between $100$ and $10'000$ times and report the median elapsed time as well as the interquartile range (IQR) of the replicates.  
Naturally, timing measurements are platform dependent. 
We produced the presented results on Intel Xeon CPU E7-2850 2.00 GHz processors using a \BIT Linux environment where \R was installed with default installation flags. 
When not indicated differently, the measurements were produced using a single thread. 

First, we consider the situation in which a pointer to an \R vector of length one is passed to the compiled C function \code{BENCHMARK()}.
The following truncated \R code illustrates how the measurements were performed.
The complete \R scripts implementing all presented performance measurements are available in the \file{benchmark} directory in the source code of \pkg{dotCall64}. 
\begin{example} 
> library("microbenchmark")
> int <- integer(1)
> microbenchmark(
+   .C("BENCHMARK", a = int, NAOK = FALSE, PACKAGE = "dotCall64"),
+   .C64("BENCHMARK", SIGNATURE = "integer", a = int, INTENT = "rw", 
+        NAOK = FALSE, PACKAGE = "dotCall64", VERBOSE = 0),
+   .C64("BENCHMARK", SIGNATURE = "integer", a = int, INTENT = "r",  
+        NAOK = FALSE, PACKAGE = "dotCall64", VERBOSE = 0),
...
\end{example}
Since the \R vector \samp{int} is very short, 
a large part of the elapsed time in this experiment is caused by the overhead of the interfaces. 
Table~\ref{tab:speedsmall} presents the resulting timing measurements in microseconds. 
They indicate that \code{.C()} is more than two times faster compared to \dc. 
However, this is not surprising, since \dc is more flexible and therefore has a larger overhead. 
The arguments \code{NAOK} and \code{INTENT} have little influence on the elapsed times.
The IQRs of around one microsecond indicate a relatively large variability of the elapsed time, which is typical for short timing measurements.

\begin{table}[H]
\centering
{\small
\begin{tabular}{lrrrrrr}
\toprule
&   \multicolumn{3}{c}{\code{NAOK = FALSE}}&\multicolumn{3}{c}{\code{NAOK = TRUE}}  \\
\cmidrule(r){2-4} \cmidrule(r){5-7}  
           & .C \phantom{000}      & .C64 [rw] \phantom{}  & .C64 [r] \phantom{0}   & .C \phantom{000}        & .C64 [rw] \phantom{}  & .C64 [r] \phantom{0} \\ 
  \midrule
double   & 2.43 (0.46) & 7.11 (0.37) & 6.97 (0.40) & 2.40 (0.45) & 7.04 (0.35) & 6.92 (0.37) \\ 
integer  & 2.39 (0.33) & 7.54 (0.85) & 7.43 (0.85) & 2.39 (0.34) & 7.52 (0.84) & 7.39 (0.83) \\ 
\BITint      &             & 8.98 (1.14) & 8.63 (1.19) &             & 8.91 (1.17) & 8.58 (1.17) \\ 
\bottomrule
\end{tabular}
}
\caption{Elapsed times in microseconds to pass double, integer, and \BITint pointers to vectors of length one from \R to C using \code{.C()} and \dc. 
The used \code{INTENT} arguments of \dc are indicated in brackets. 
Reported are median elapsed times of $10'000$ replicates.
The corresponding IQRs are indicated in parentheses. 
}
\label{tab:speedsmall}
\end{table}

We repeated the same experiment with vectors of length $2^{28}$. 
Now, the elapsed times are dominated by services of the interfaces (\ie, checking for missing\slash infinite values, copying, and casting).
The timings in seconds are presented in Table~\ref{tab:speedsmall}. 
They indicate that \dc with argument \code{INTENT} \code{= "rw"} and \code{.C()} showed similar elapsed times.
When the intent is set to ``read'' (\code{INTENT} \code{= "r"}), the elapsed times were reduced and dropped to $0.00$ seconds for some configurations. 
Moreover, not checking for missing\slash infinite values (\code{NAOK} \code{= TRUE}) decreases the elapsed times across all considered cases.
The castings of \code{SIGNATURE} \code{= "int64"} arguments seems to be the most time-consuming task. 
Note that the IQRs are now smaller relative to the measured timings, because the measured times are larger.

\begin{table}[H]
\centering
{\small
\begin{tabular}{lrrrrrr}
\toprule
&   \multicolumn{3}{c}{\code{NAOK = FALSE}}&\multicolumn{3}{c}{\code{NAOK = TRUE}}  \\
\cmidrule(r){2-4} \cmidrule(r){5-7}  
           & .C \phantom{000}      & .C64 [rw] \phantom{}  & .C64 [r] \phantom{0}   & .C \phantom{000}        & .C64 [rw] \phantom{}  & .C64 [r] \phantom{0} \\ 
  \midrule
double  & 2.65 (0.05) & 3.16 (0.06) & 1.82 (0.02) & 1.33 (0.06) & 1.33 (0.05) & 0.00 (0.00) \\ 
integer & 1.09 (0.03) & 1.09 (0.04) & 0.43 (0.01) & 0.66 (0.03) & 0.66 (0.04) & 0.00 (0.00) \\ 
\BITint     &             & 5.21 (0.20) & 3.80 (0.06) &             & 3.36 (0.06) & 1.97 (0.06) \\ 
\bottomrule
\end{tabular}
}
\caption{Elapsed times in seconds to pass double, integer, and \BITint pointers to vectors of length $2^{28}$ from \R\ to C using \code{.C()} and \dc. 
The used \code{INTENT} arguments of \dc are indicated in brackets. 
Reported are median elapsed times of $100$ replicates.
The corresponding IQRs are indicated in parentheses. 
}
\label{tab:speed}
\end{table}

In a second series of timing measurements, we consider the situation in which a pointer to a vector is passed to the compiled code to write into the vector.
We measure the elapsed times of this task as shown in the following truncated \R code.
\begin{example}
> microbenchmark(
+   .C("BENCHMARK", a = integer(2^28), NAOK = TRUE, package = "dotCall64")
+   .C64("BENCHMARK", SIGNATURE = "integer", a = integer(2^28), INTENT = "rw", 
+        NAOK = TRUE, package = "dotCall64", VERBOSE = 0)
+   .C64("BENCHMARK", SIGNATURE = "integer", a = integer_dc(2^28), INTENT = "w", 
+        NAOK = TRUE, package = "dotCall64", VERBOSE = 0),
...
\end{example}
Note the usage of \code{integer\_dc()}, which creates a list containing the length and class of the vector.
This information is then used by \dc to create the corresponding vector in C.
Table~\ref{tab:speed2} shows the timing measurements for the described setting.
As expected using \dc with \code{INTENT} \code{= "w"} reduces the elapsed times 
compared to \code{INTENT} \code{= "rw"} substantially. 
Furthermore, \code{.C()} and  \dc with \code{INTENT} \code{= "w"} have similar elapsed times.
While \code{.C()} relies on the reference counting mechanism of \R objects to avoid copying \citep[``Writing \R Extensions,''][]{r_ext}, \dc uses the \code{"vector\_dc"} class.
The latter has the advantage that one \double to \BITint casting can be avoided in the \code{SIGNATURE} \code{= "int64"} case. 

\begin{table}[H]
\centering
{\small
\begin{tabular}{lrrr}
\toprule
 & .C \phantom{000.}& .C64 [rw] \phantom{.}& .C64 [w] \phantom{0}\\ 
  \midrule
double  & 0.87 (0.01) & 2.28 (0.13) & 0.87 (0.01) \\ 
integer & 0.44 (0.01) & 1.16 (0.06) & 0.44 (0.01) \\ 
\BITint     &             & 4.27 (0.03) & 2.27 (0.02) \\ 
   \bottomrule
\end{tabular}
}
\caption{Elapsed times in seconds to pass double, integer, and \BITint pointers to vectors of length $2^{28}$ initialized with zeros from \R to C using \code{.C()} and \dc. 
The used \code{INTENT} arguments of \dc are indicated in brackets. 
Reported are median elapsed times of $100$ replicates.
The corresponding IQRs are indicated in parentheses. 
}
\label{tab:speed2}
\end{table}

\begin{figure}[tb]
\begin{center}
\includegraphics[trim = 5mm 0mm 0mm 0mm, clip, height=.37\textwidth]{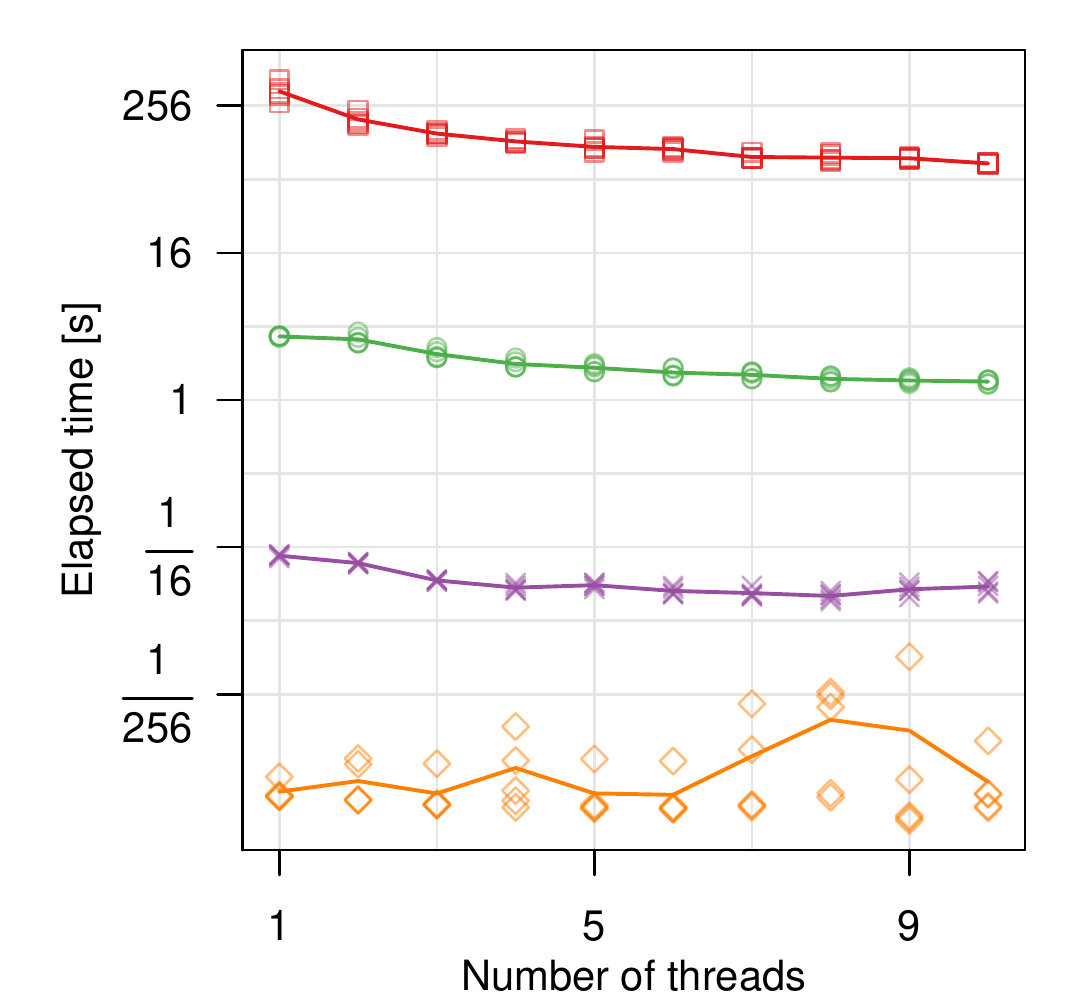} 
\hspace*{5mm}
\includegraphics[trim = 0mm 0mm 0mm 0mm, clip, height=.37\textwidth]{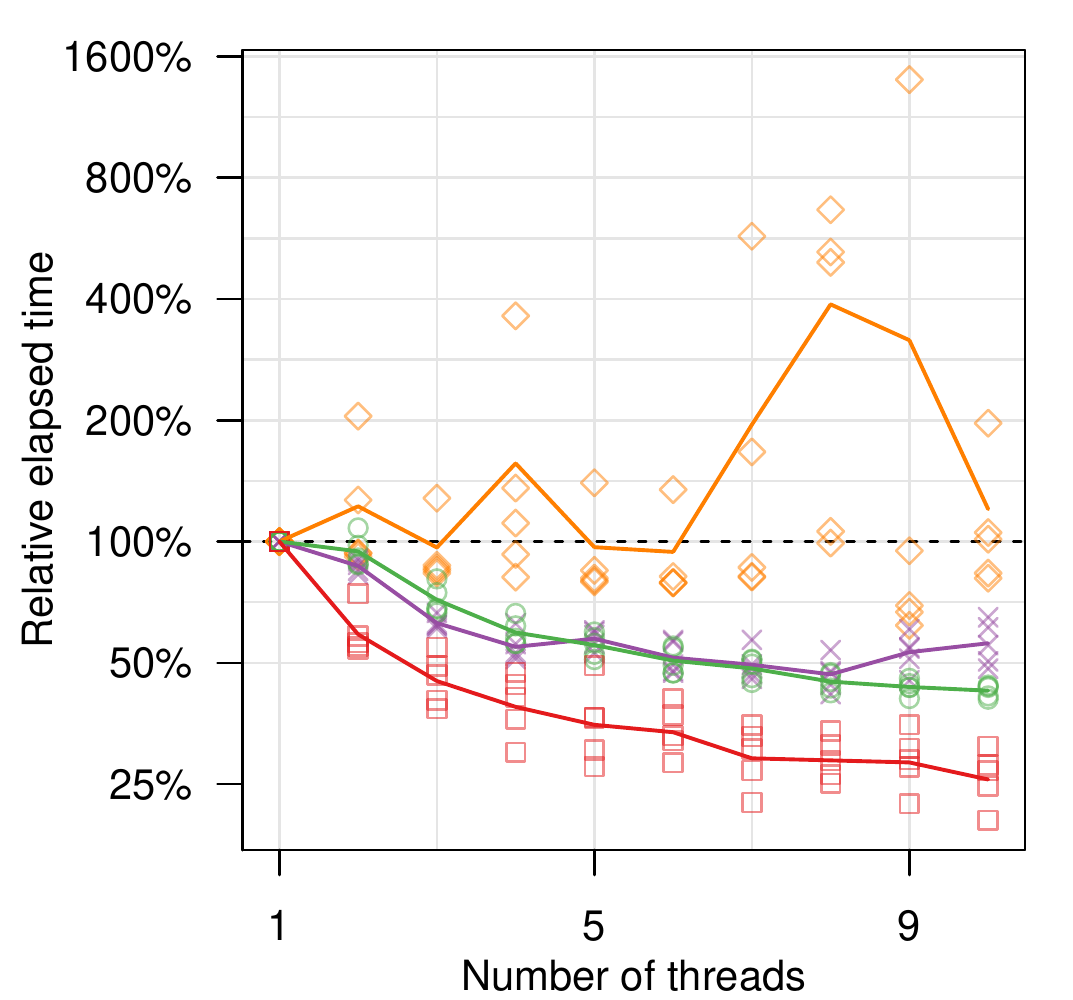} 
\hspace*{1mm}
\raisebox{1cm}{\includegraphics[trim = 80mm 10mm 55mm 30mm, clip, width=.15\textwidth]{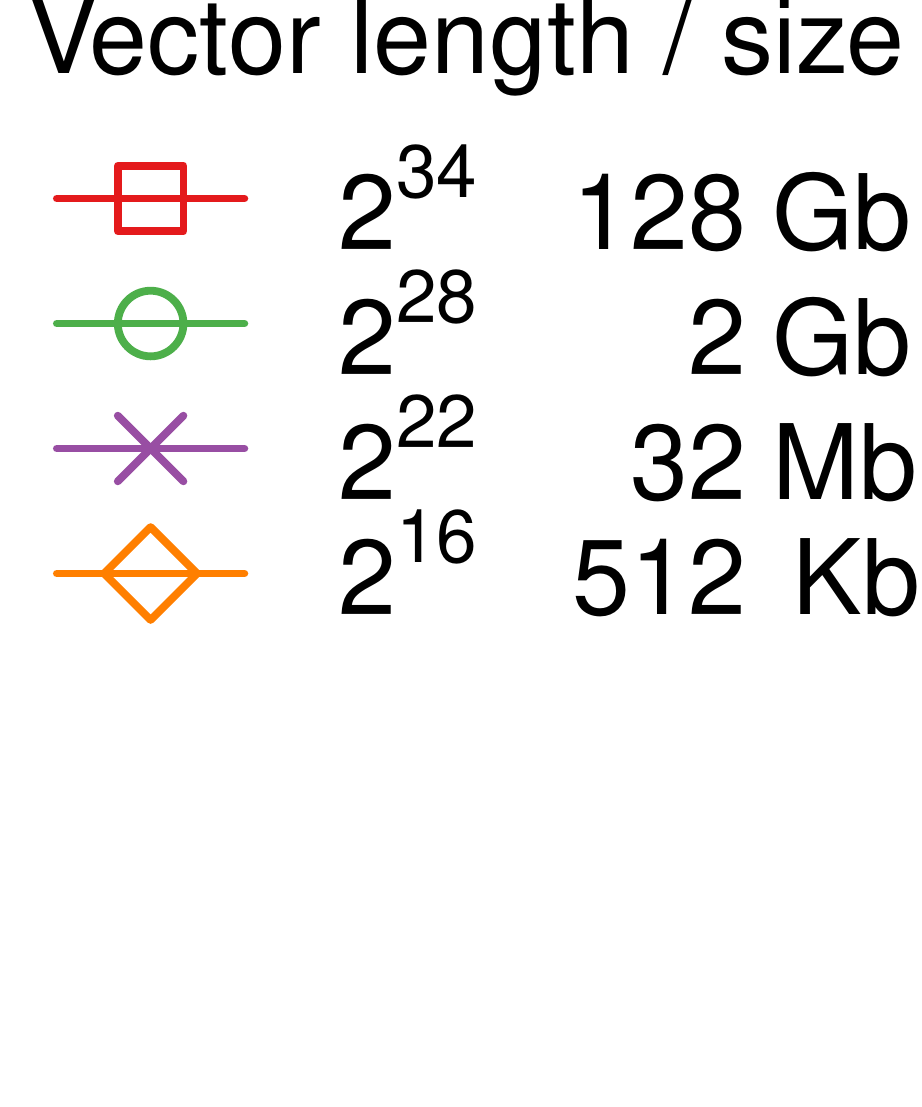}} 
\end{center}
\caption{
Timings measurements to illustrate the effect of using \dc with enabled multithreading (openMP).
Colors and symbols indicate the length\slash size of the evaluated vectors.
Five replicates of each measured configuration are shown with symbols, and the mean values thereof are connected with a line.
Left panel: The elapsed time in seconds ($y$-axis) is plotted against the number of used threads ($x$-axis).  
Right panel: The decrease\slash increase in elapsed time relative to using one thread ($y$-axis) is plotted 
against the number of threads ($x$-axis).
}
\label{fig:openMP}
\end{figure}

The function \dc features an openMP implementation of the \double to \BITint and \BITint to \double castings of
\code{SIGNATURE} \code{="int64"} arguments.
Hence, the computational workload of the castings can be distributed to several threads running in parallel.
To quantify the performance gain related to using openMP, 
we control the number of used threads to be between $1$ and $10$ with the \R~package \pkg{OpenMPController} and measure the elapsed times of the following call.
\begin{example} 
> .C64("BENCHMARK", SIGNATURE = "int64", a = a, INTENT = "rw", NAOK = TRUE,  
+      PACKAGE = "dotCall64", VERBOSE = 0)
\end{example}
We let \samp{a} be \double vectors of length $2^{16}$, $2^{22}$, $2^{28}$, and $2^{34}$ and performed five replicated timing measurements for each configuration.
The results are summarized in Figure~\ref{fig:openMP}.
The reduction in computation time due to using multiple threads is greatest for the vectors of length $2^{34}$, where using $10$ threads reduced the elapsed times by about $70\%$.
Conversely, for the vector of length $2^{16}$ no reduction was observed.

\section{Summary}
This paper presents the \R~package \pkg{dotCall64}, which provides an alternative to \code{.C()} and \code{.Fortran()} from the \ffi. 
In the first section, we introduce \R's interfaces to embed compiled \cpp and Fortran code. 
We argue that, in some situations, a \code{.C()} type interface is more convenient compared to using the C API of \R in conjunction with the \mic. 
In section two, we motivate the development of \pkg{dotCall64} with a discussion of missing features of the \ffi and an overview of the \R implementation of \longvectors.
Then, we present the usage and the implementation of the \dc function from the \R package \pkg{dotCall64}.
This is followed by examples demonstrating the capabilities of the new interface---also in comparison with the \ffi. 
Furthermore, we discuss strategies to extend entire \R~packages with compiled code supporting \longvectors. 
In the last section, we present performance measurements of the \dc interface and benchmark it against \code{.C()}. 
This highlights the speed gains achieved by avoiding unnecessary copies of \R vectors and by using openMP for casting \R vectors. 
In conclusion, the interface provided by the \R~package \pkg{dotCall64} is an up-to-date version of the \ffi including tools to conveniently embed compiled code manipulating \longvectors. 

\section{Acknowledgments}
We thank Rafael Ostertag for contributions to Figure~\ref{fig:venn}. 
We acknowledge the support of the University of Zurich Research Priority Program (URPP) on “Global Change and Biodiversity.”

\bibliography{Gerber_dotCall64}

\address{Florian Gerber\\
Department of Mathematics\\
University of Zurich\\
8057 Zurich\\ 
Switzerland\\}
\email{florian.gerber@math.uzh.ch}

\address{Kaspar M\"{o}singer\\
Department of Mathematics\\
University of Zurich\\
8057 Zurich\\ 
Switzerland\\}
\email{kaspar.moesinger@gmail.com}

\address{Reinhard Furrer\\
Department of Mathematics and \\
Department of Computational Science\\
University of Zurich\\
8057 Zurich\\ 
Switzerland\\}
\email{reinhard.furrer@math.uzh.ch}

\section{Appendix: R source code}
In the following, we show parts of the C source code of \R version~3.3.1 to support the understanding of the \longvector implementation. 
More precisely, the lines 26--377 from the file \file{Rinternals.h} and the lines 124--191 from the file \file{Rinlinedfuns.h} are shown in Listing~\ref{list:Rinternals} and Listing~\ref{list:Rinlinedfuns}, respectively. 
The indicated line numbers in the code refer to the actual line numbers of the corresponding file.  

\renewcommand{\thelstlisting}{\arabic{lstlisting}}
\begin{lstlisting}[firstnumber=26,firstline=26,lastline=377,label=list:Rinternals,caption=\file{R-3.3.1/src/include/Rinternals.h}]
#ifndef R_INTERNALS_H_
#define R_INTERNALS_H_

// Support for NO_C_HEADERS added in R 3.3.0
#ifdef __cplusplus
# ifndef NO_C_HEADERS
#  include <cstdio>
#  ifdef __SUNPRO_CC
using std::FILE;
#  endif
#  include <climits>
#  include <cstddef>
# endif
extern "C" {
#else
# ifndef NO_C_HEADERS
#  include <stdio.h>
#  include <limits.h> /* for INT_MAX */
#  include <stddef.h> /* for ptrdiff_t */
# endif
#endif

#include <R_ext/Arith.h>
#include <R_ext/Boolean.h>
#include <R_ext/Complex.h>
#include <R_ext/Error.h>  // includes NORET macro
#include <R_ext/Memory.h>
#include <R_ext/Utils.h>
#include <R_ext/Print.h>

#include <R_ext/libextern.h>

typedef unsigned char Rbyte;

/* type for length of (standard, not long) vectors etc */
typedef int R_len_t;
#define R_LEN_T_MAX INT_MAX

/* both config.h and Rconfig.h set SIZEOF_SIZE_T, but Rconfig.h is
   skipped if config.h has already been included. */
#ifndef R_CONFIG_H
# include <Rconfig.h>
#endif

#if ( SIZEOF_SIZE_T > 4 )
# define LONG_VECTOR_SUPPORT
#endif

#ifdef LONG_VECTOR_SUPPORT
    typedef ptrdiff_t R_xlen_t;
    typedef struct { R_xlen_t lv_length, lv_truelength; } R_long_vec_hdr_t;
# define R_XLEN_T_MAX 4503599627370496
# define R_SHORT_LEN_MAX 2147483647
# define R_LONG_VEC_TOKEN -1
#else
    typedef int R_xlen_t;
# define R_XLEN_T_MAX R_LEN_T_MAX
#endif

#ifndef TESTING_WRITE_BARRIER
# define INLINE_PROTECT
#endif

/* Fundamental Data Types:  These are largely Lisp
 * influenced structures, with the exception of LGLSXP,
 * INTSXP, REALSXP, CPLXSXP and STRSXP which are the
 * element types for S-like data objects.
 *
 *			--> TypeTable[] in ../main/util.c for  typeof()
 */

/*  These exact numeric values are seldom used, but they are, e.g., in
 *  ../main/subassign.c, and they are serialized.
*/
#ifndef enum_SEXPTYPE
/* NOT YET using enum:
 *  1)	The SEXPREC struct below has 'SEXPTYPE type : 5'
 *	(making FUNSXP and CLOSXP equivalent in there),
 *	giving (-Wall only ?) warnings all over the place
 * 2)	Many switch(type) { case ... } statements need a final `default:'
 *	added in order to avoid warnings like [e.g. l.170 of ../main/util.c]
 *	  "enumeration value `FUNSXP' not handled in switch"
 */
typedef unsigned int SEXPTYPE;

#define NILSXP	     0	  /* nil = NULL */
#define SYMSXP	     1	  /* symbols */
#define LISTSXP	     2	  /* lists of dotted pairs */
#define CLOSXP	     3	  /* closures */
#define ENVSXP	     4	  /* environments */
#define PROMSXP	     5	  /* promises: [un]evaluated closure arguments */
#define LANGSXP	     6	  /* language constructs (special lists) */
#define SPECIALSXP   7	  /* special forms */
#define BUILTINSXP   8	  /* builtin non-special forms */
#define CHARSXP	     9	  /* "scalar" string type (internal only)*/
#define LGLSXP	    10	  /* logical vectors */
/* 11 and 12 were factors and ordered factors in the 1990s */
#define INTSXP	    13	  /* integer vectors */
#define REALSXP	    14	  /* real variables */
#define CPLXSXP	    15	  /* complex variables */
#define STRSXP	    16	  /* string vectors */
#define DOTSXP	    17	  /* dot-dot-dot object */
#define ANYSXP	    18	  /* make "any" args work.
			     Used in specifying types for symbol
			     registration to mean anything is okay  */
#define VECSXP	    19	  /* generic vectors */
#define EXPRSXP	    20	  /* expressions vectors */
#define BCODESXP    21    /* byte code */
#define EXTPTRSXP   22    /* external pointer */
#define WEAKREFSXP  23    /* weak reference */
#define RAWSXP      24    /* raw bytes */
#define S4SXP       25    /* S4, non-vector */

/* used for detecting PROTECT issues in memory.c */
#define NEWSXP      30    /* fresh node created in new page */
#define FREESXP     31    /* node released by GC */

#define FUNSXP      99    /* Closure or Builtin or Special */


#else /* NOT YET */
/*------ enum_SEXPTYPE ----- */
typedef enum {
    NILSXP	= 0,	/* nil = NULL */
    SYMSXP	= 1,	/* symbols */
    LISTSXP	= 2,	/* lists of dotted pairs */
    CLOSXP	= 3,	/* closures */
    ENVSXP	= 4,	/* environments */
    PROMSXP	= 5,	/* promises: [un]evaluated closure arguments */
    LANGSXP	= 6,	/* language constructs (special lists) */
    SPECIALSXP	= 7,	/* special forms */
    BUILTINSXP	= 8,	/* builtin non-special forms */
    CHARSXP	= 9,	/* "scalar" string type (internal only)*/
    LGLSXP	= 10,	/* logical vectors */
    INTSXP	= 13,	/* integer vectors */
    REALSXP	= 14,	/* real variables */
    CPLXSXP	= 15,	/* complex variables */
    STRSXP	= 16,	/* string vectors */
    DOTSXP	= 17,	/* dot-dot-dot object */
    ANYSXP	= 18,	/* make "any" args work */
    VECSXP	= 19,	/* generic vectors */
    EXPRSXP	= 20,	/* expressions vectors */
    BCODESXP	= 21,	/* byte code */
    EXTPTRSXP	= 22,	/* external pointer */
    WEAKREFSXP	= 23,	/* weak reference */
    RAWSXP	= 24,	/* raw bytes */
    S4SXP	= 25,	/* S4 non-vector */

    NEWSXP      = 30,   /* fresh node creaed in new page */
    FREESXP     = 31,   /* node released by GC */

    FUNSXP	= 99	/* Closure or Builtin */
} SEXPTYPE;
#endif

/* These are also used with the write barrier on, in attrib.c and util.c */
#define TYPE_BITS 5
#define MAX_NUM_SEXPTYPE (1<<TYPE_BITS)

// ======================= USE_RINTERNALS section
#ifdef USE_RINTERNALS
/* This is intended for use only within R itself.
 * It defines internal structures that are otherwise only accessible
 * via SEXP, and macros to replace many (but not all) of accessor functions
 * (which are always defined).
 */

/* Flags */


struct sxpinfo_struct {
    SEXPTYPE type      :  TYPE_BITS;/* ==> (FUNSXP == 99) %% 2^5 == 3 == CLOSXP
			     * -> warning: `type' is narrower than values
			     *              of its type
			     * when SEXPTYPE was an enum */
    unsigned int obj   :  1;
    unsigned int named :  2;
    unsigned int gp    : 16;
    unsigned int mark  :  1;
    unsigned int debug :  1;
    unsigned int trace :  1;  /* functions and memory tracing */
    unsigned int spare :  1;  /* currently unused */
    unsigned int gcgen :  1;  /* old generation number */
    unsigned int gccls :  3;  /* node class */
}; /*		    Tot: 32 */

struct vecsxp_struct {
    R_len_t	length;
    R_len_t	truelength;
};

struct primsxp_struct {
    int offset;
};

struct symsxp_struct {
    struct SEXPREC *pname;
    struct SEXPREC *value;
    struct SEXPREC *internal;
};

struct listsxp_struct {
    struct SEXPREC *carval;
    struct SEXPREC *cdrval;
    struct SEXPREC *tagval;
};

struct envsxp_struct {
    struct SEXPREC *frame;
    struct SEXPREC *enclos;
    struct SEXPREC *hashtab;
};

struct closxp_struct {
    struct SEXPREC *formals;
    struct SEXPREC *body;
    struct SEXPREC *env;
};

struct promsxp_struct {
    struct SEXPREC *value;
    struct SEXPREC *expr;
    struct SEXPREC *env;
};

/* Every node must start with a set of sxpinfo flags and an attribute
   field. Under the generational collector these are followed by the
   fields used to maintain the collector's linked list structures. */

/* Define SWITH_TO_REFCNT to use reference counting instead of the
   'NAMED' mechanism. This uses the R-devel binary layout. The two
   'named' field bits are used for the REFCNT, so REFCNTMAX is 3. */
//#define SWITCH_TO_REFCNT

#if defined(SWITCH_TO_REFCNT) && ! defined(COMPUTE_REFCNT_VALUES)
# define COMPUTE_REFCNT_VALUES
#endif
#define REFCNTMAX (4 - 1)

#define SEXPREC_HEADER \
    struct sxpinfo_struct sxpinfo; \
    struct SEXPREC *attrib; \
    struct SEXPREC *gengc_next_node, *gengc_prev_node

/* The standard node structure consists of a header followed by the
   node data. */
typedef struct SEXPREC {
    SEXPREC_HEADER;
    union {
	struct primsxp_struct primsxp;
	struct symsxp_struct symsxp;
	struct listsxp_struct listsxp;
	struct envsxp_struct envsxp;
	struct closxp_struct closxp;
	struct promsxp_struct promsxp;
    } u;
} SEXPREC, *SEXP;

/* The generational collector uses a reduced version of SEXPREC as a
   header in vector nodes.  The layout MUST be kept consistent with
   the SEXPREC definition.  The standard SEXPREC takes up 7 words on
   most hardware; this reduced version should take up only 6 words.
   In addition to slightly reducing memory use, this can lead to more
   favorable data alignment on 32-bit architectures like the Intel
   Pentium III where odd word alignment of doubles is allowed but much
   less efficient than even word alignment. */
typedef struct VECTOR_SEXPREC {
    SEXPREC_HEADER;
    struct vecsxp_struct vecsxp;
} VECTOR_SEXPREC, *VECSEXP;

typedef union { VECTOR_SEXPREC s; double align; } SEXPREC_ALIGN;

/* General Cons Cell Attributes */
#define ATTRIB(x)	((x)->attrib)
#define OBJECT(x)	((x)->sxpinfo.obj)
#define MARK(x)		((x)->sxpinfo.mark)
#define TYPEOF(x)	((x)->sxpinfo.type)
#define NAMED(x)	((x)->sxpinfo.named)
#define RTRACE(x)	((x)->sxpinfo.trace)
#define LEVELS(x)	((x)->sxpinfo.gp)
#define SET_OBJECT(x,v)	(((x)->sxpinfo.obj)=(v))
#define SET_TYPEOF(x,v)	(((x)->sxpinfo.type)=(v))
#define SET_NAMED(x,v)	(((x)->sxpinfo.named)=(v))
#define SET_RTRACE(x,v)	(((x)->sxpinfo.trace)=(v))
#define SETLEVELS(x,v)	(((x)->sxpinfo.gp)=((unsigned short)v))

#if defined(COMPUTE_REFCNT_VALUES)
# define REFCNT(x) ((x)->sxpinfo.named)
# define TRACKREFS(x) (TYPEOF(x) == CLOSXP ? TRUE : ! (x)->sxpinfo.spare)
#else
# define REFCNT(x) 0
# define TRACKREFS(x) FALSE
#endif

#ifdef SWITCH_TO_REFCNT
# undef NAMED
# undef SET_NAMED
# define NAMED(x) REFCNT(x)
# define SET_NAMED(x, v) do {} while (0)
#endif

/* S4 object bit, set by R_do_new_object for all new() calls */
#define S4_OBJECT_MASK ((unsigned short)(1<<4))
#define IS_S4_OBJECT(x) ((x)->sxpinfo.gp & S4_OBJECT_MASK)
#define SET_S4_OBJECT(x) (((x)->sxpinfo.gp) |= S4_OBJECT_MASK)
#define UNSET_S4_OBJECT(x) (((x)->sxpinfo.gp) &= ~S4_OBJECT_MASK)

/* Vector Access Macros */
#ifdef LONG_VECTOR_SUPPORT
    R_len_t NORET R_BadLongVector(SEXP, const char *, int);
# define IS_LONG_VEC(x) (SHORT_VEC_LENGTH(x) == R_LONG_VEC_TOKEN)
# define SHORT_VEC_LENGTH(x) (((VECSEXP) (x))->vecsxp.length)
# define SHORT_VEC_TRUELENGTH(x) (((VECSEXP) (x))->vecsxp.truelength)
# define LONG_VEC_LENGTH(x) ((R_long_vec_hdr_t *) (x))[-1].lv_length
# define LONG_VEC_TRUELENGTH(x) ((R_long_vec_hdr_t *) (x))[-1].lv_truelength
# define XLENGTH(x) (IS_LONG_VEC(x) ? LONG_VEC_LENGTH(x) : SHORT_VEC_LENGTH(x))
# define XTRUELENGTH(x)	(IS_LONG_VEC(x) ? LONG_VEC_TRUELENGTH(x) : SHORT_VEC_TRUELENGTH(x))
# define LENGTH(x) (IS_LONG_VEC(x) ? R_BadLongVector(x, __FILE__, __LINE__) : SHORT_VEC_LENGTH(x))
# define TRUELENGTH(x) (IS_LONG_VEC(x) ? R_BadLongVector(x, __FILE__, __LINE__) : SHORT_VEC_TRUELENGTH(x))
# define SET_SHORT_VEC_LENGTH(x,v) (SHORT_VEC_LENGTH(x) = (v))
# define SET_SHORT_VEC_TRUELENGTH(x,v) (SHORT_VEC_TRUELENGTH(x) = (v))
# define SET_LONG_VEC_LENGTH(x,v) (LONG_VEC_LENGTH(x) = (v))
# define SET_LONG_VEC_TRUELENGTH(x,v) (LONG_VEC_TRUELENGTH(x) = (v))
# define SETLENGTH(x,v) do { \
      SEXP sl__x__ = (x); \
      R_xlen_t sl__v__ = (v); \
      if (IS_LONG_VEC(sl__x__)) \
	  SET_LONG_VEC_LENGTH(sl__x__,  sl__v__); \
      else SET_SHORT_VEC_LENGTH(sl__x__, (R_len_t) sl__v__); \
  } while (0)
# define SET_TRUELENGTH(x,v) do { \
      SEXP sl__x__ = (x); \
      R_xlen_t sl__v__ = (v); \
      if (IS_LONG_VEC(sl__x__)) \
	  SET_LONG_VEC_TRUELENGTH(sl__x__, sl__v__); \
      else SET_SHORT_VEC_TRUELENGTH(sl__x__, (R_len_t) sl__v__); \
  } while (0)
# define IS_SCALAR(x, type) (TYPEOF(x) == (type) && SHORT_VEC_LENGTH(x) == 1)
#else
# define SHORT_VEC_LENGTH(x) (((VECSEXP) (x))->vecsxp.length)
# define LENGTH(x)	(((VECSEXP) (x))->vecsxp.length)
# define TRUELENGTH(x)	(((VECSEXP) (x))->vecsxp.truelength)
# define XLENGTH(x) LENGTH(x)
# define XTRUELENGTH(x) TRUELENGTH(x)
# define SETLENGTH(x,v)		((((VECSEXP) (x))->vecsxp.length)=(v))
# define SET_TRUELENGTH(x,v)	((((VECSEXP) (x))->vecsxp.truelength)=(v))
# define SET_SHORT_VEC_LENGTH SETLENGTH
# define SET_SHORT_VEC_TRUELENGTH SET_TRUELENGTH
# define IS_LONG_VEC(x) 0
# define IS_SCALAR(x, type) (TYPEOF(x) == (type) && LENGTH(x) == 1)
#endif
\end{lstlisting}

\begin{lstlisting}[firstnumber=124,firstline=124,lastline=191,label=list:Rinlinedfuns,caption=\file{R-3.3.1/src/include/Rinlinedfuns.h}]
INLINE_FUN R_len_t length(SEXP s)
{
    switch (TYPEOF(s)) {
    case NILSXP:
	return 0;
    case LGLSXP:
    case INTSXP:
    case REALSXP:
    case CPLXSXP:
    case STRSXP:
    case CHARSXP:
    case VECSXP:
    case EXPRSXP:
    case RAWSXP:
	return LENGTH(s);
    case LISTSXP:
    case LANGSXP:
    case DOTSXP:
    {
	int i = 0;
	while (s != NULL && s != R_NilValue) {
	    i++;
	    s = CDR(s);
	}
	return i;
    }
    case ENVSXP:
	return Rf_envlength(s);
    default:
	return 1;
    }
}

R_xlen_t Rf_envxlength(SEXP rho);

INLINE_FUN R_xlen_t xlength(SEXP s)
{
    switch (TYPEOF(s)) {
    case NILSXP:
	return 0;
    case LGLSXP:
    case INTSXP:
    case REALSXP:
    case CPLXSXP:
    case STRSXP:
    case CHARSXP:
    case VECSXP:
    case EXPRSXP:
    case RAWSXP:
	return XLENGTH(s);
    case LISTSXP:
    case LANGSXP:
    case DOTSXP:
    {
	// it is implausible this would be >= 2^31 elements, but allow it
	R_xlen_t i = 0;
	while (s != NULL && s != R_NilValue) {
	    i++;
	    s = CDR(s);
	}
	return i;
    }
    case ENVSXP:
	return Rf_envxlength(s);
    default:
	return 1;
    }
}
\end{lstlisting}

\end{article}

\end{document}